# Non-Thermal Mechanism of Weak Microwave Fields Influence on Nerve Fiber

M.N. Shneider[1*] and M. Pekker[2]

[1]Department of Mechanical and Aerospace Engineering, Princeton University, Princeton, NJ 08544, USA

[2]Drexel Plasma Institute, Drexel University, 200 Federal Street, Camden, NJ 08103, USA

E-mail: m.n.shneider@gmail.com

**Abstract**

We propose a non-thermal mechanism of weak microwave field impact on a nerve fiber. It is shown that in the range of about 30 - 300 GHz there are strongly pronounced resonances associated with the excitation of ultrasonic vibrations in the membrane as a result of interaction with electromagnetic radiation. These vibrations create acoustic pressure which may lead to the redistribution of the protein transmembrane channels, and, thus, changing the threshold of the action potential excitation in the axons of the neural network. The influence of the electromagnetic microwave radiation on various specific areas of myelin nerve fibers was analyzed: the nodes of Ranvier, and the so-called initial segment - the area between the neuron hillock and the first part of the axon covered with the myelin layer. It is shown that the initial segment is the most sensitive area of the myelined nerve fibers from which the action potential normally starts.

## 1. Introduction

Investigations of the interaction of non-thermal low intensity microwave field with cell membranes and cellular structures began in the 60's and 70's of last century. Early experiments exploring physiological effects of non-thermal microwave field intensity, and their possible theoretical explanation, are reviewed in the monograph [Devyatkov et al, 1991]. However, studies of non-thermal effect of microwave radiation on a nerve and nerve activity were not carried out until recently. In the past decades, the level of electromagnetic pollution of the human environment has grown dramatically. This is in large part is related to the current state of electronic communications, e.g. cellular phone networks, radio and television, WiFi, etc. In recent years it has been proposed to increase the frequency of the electromagnetic radiation used for telecommunications up to 30-300 GHz [Marcus and Pattan, 2005; Lawton, 2008]. This is based on the fact that at such frequencies it is possible to transmit signals at a rate higher than 2 gigabits per second, which has enormous potential both for civilian purposes, as well as for the military. On the other hand, the increase in frequency makes it possible to use low-power energy sources that essentially reduce the electromagnetic radiation flux down to lower than 1

---

[*] Author to whom correspondence should be addressed

W/m$^2$. At such power of the incident radiation, the thermal effect on the nerve fiber is insignificant and cannot affect the vital functions of the human brain and the nervous system. However, recently, several experimental studies had been done, for example, see [Pikov et al, 2010], which showed that a relatively weak microwave radiation with intensities of 0.01 - 1 W/m$^2$ in the range of 50 -100 GHz leads to the spontaneous excitation of neural activity. These results cannot be explained within the standard thermal mechanism. Note, in the papers [Pakhomov et al 1998, Rojavin, Ziskin, 1998], in particular, it was pointed out that the known therapeutic effect on humans by the weak sources of microwave radiation ~ 10 W/m$^2$, operating at frequencies 42.25, 53.57, or 61.22 GHz, also cannot be explained by heating.

If we look at [IEEE Standard, 2005] in which the acceptable norms of radiation fluxes for each microwave frequency band are given, it is seen that the permissible rate of radiation calculated in the range of 3-300 GHz on the basis of a thermal effect is 100 W/m$^2$, which is approximately ten thousand times higher than the microwave intensity levels at which the spontaneous excitation of the nerve fibers was observed in the experiments [Pikov et al, 2010]. Therefore, the studies on dosimetry and the analysis of microwave radiation acting on living organisms are very important [Zhadobov et al, 2008].

In this paper we propose a non-thermal mechanism of action of weak microwave fields on the nerve fiber. It is shown that in a relatively weak microwave frequency in the frequency range of about 30 - 300 GHz there are pronounced resonances associated with the excitation of ultrasonic vibrations in the membrane which may lead to redistribution of the transmembrane protein ion channels. The characteristic resonance frequencies are found, and it is shown that the impact of a low-intensity microwave radiation of order 0.1 - 10 W/m$^2$ is already enough for redistribution of protein ion channels which may impact the value of the threshold for the action potentials initiation in the neural network. The influence of the electromagnetic microwave radiation on different characteristic parts of the nerve fibers was analyzed: the nodes of Ranvier and the so-called initial segment - the area between the neuron hillock and the first part of an axon covered with the myelin layer. It is shown that the initial segment is the most sensitive area of myelined nerve fibers from which the action potential normally starts.

**2. Transmembrane protein channels displacement under the action of ultrasound**

Ultrasound is widely used in medicine for diagnostics, as well as, for anesthesia [Robertson and Baker,2001; Foley et al, 2008; Colucci et al, 2009]. Thus, depending on the frequency, amplitude and time of exposure, pain relief occurs due to the local heating caused by the ultrasound absorption in a reversible [Foley et al, 2008; Colucci et al, 2009; Young and Henneman, 1961] as well as irreversible manner (see for example, [Vykhodtseva et al, 1976]) due to the irreversible destruction of the nerve fibers in the high intensity ultrasonic wave.

We will show that, with the increasing ultrasound frequency, a non-thermal mechanism of excitation or inhibition of the conditions for the formation and transmission of the action potential can become dominant. It is known that the transport of sodium ions through a bilayer lipid membrane is carried out through specific protein channels (Fig. 1). Under normal conditions, these channels are distributed uniformly within the membrane. Leveling of the surface density of these channels occurs due to the so-called lateral diffusion [Jacobson, 1987; Almeida and Vaz, 1995; Ramadurai et al, 2009].

There are several possibilities to change (increase or decrease) the threshold of the action potential initiation. This can be done by the local heating of the membrane (thermal mechanism) [Colucci et al, 2009] or by changing in some way the selective ion permeability of the nerve fibers. Thus, for example, as was shown in [Platkiewicz and Brette, 2010], the change in the density of ion channels within tens of percent of the equilibrium state can significantly alter the action potential threshold in comparison to the unperturbed membrane. This fact, at certain conditions, significantly increases the probability of the spontaneous formation of the action potential.

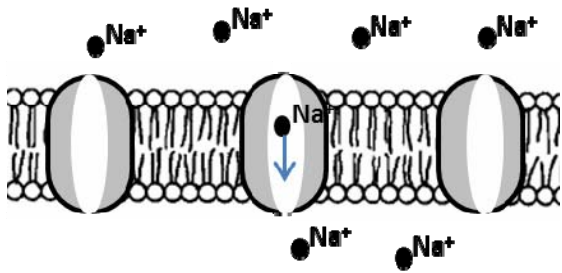

**Figure 1**. An element of the lipid bilayer membrane of the axon with $Na^+$ protein transmembrane channels

An ultrasound may serve as ssuch factor that changes the uniform distribution of the transmembrane channels. It is known that an ultrasound in a fluid with suspended particles leads to coagulation, i.e. the appearance of inhomogeneous density distribution of suspended particles (see, eg, [Mednikov, 1965]). An example of this process can be the experimentally observed concentration / separation of the suspension of living cells of different sizes and shapes [Sadikova and Pashovkin, 2013].

In this paper, we restrict the consideration of the interaction of ultrasonic waves with transmembrane proteins in the simplest approximation by considering the protein channels as incompressible spheres with the radius $a = h = d/2$, where $d$ is the thickness of membrane. The acoustic radiation force acting on the incompressible sphere of radius $a \ll \lambda$ at the traveling sound wave is [King, 1934]:

$$F_t = 4\pi a^5 k^4 \frac{1+\frac{2}{9}(1-\varsigma)^2}{(2+\varsigma)^2} \overline{E}. \tag{1}$$

Here $\overline{E} = \frac{1}{2}\rho\omega^2 A^2$ is the energy density in the wave; $\omega = 2\pi f$ is the angular frequency of ultrasound; $A$ is the amplitude of the oscillatory displacements in the wave; $k = 2\pi/\lambda = \omega/c_s$ is the wave number; $\varsigma = \rho/\rho_a$, $\rho, \rho_a$ are the densities of the medium and the particles.

In the case of a standing acoustic wave, the force acting on the suspended spherical incompressible particle is [King, 1934]:

$$F_s = 2\pi a^3 k_n \overline{E} \sin(2k_n x)[1+\tfrac{2}{3}(1-\varsigma)]/(2+\varsigma), \tag{2}$$

where $x$ is the position of a particle in the standing wave; $k_n = \frac{n\pi}{l}$, $n = 1,2,3...$, $l$ is the size of the area where a standing wave exists. In the case where the particle density is practically the same as the density of the fluid, $\varsigma \approx 1$, the force (2) has the form:

$$F_s = \frac{2\pi}{3} \cdot a^3 k_n \overline{E} \sin(2k_n x). \tag{3}$$

Under the action of acoustic radiation pressure force (1) or (2) the transmembrane proteins channels are displaced along the $x$-axis as long as the reverse diffusion flux is formed, arising due to the emergence of the channels surface density gradient. Thus, the stationary distribution of the protein channels density can be found from the condition that their total flux vanishes:

$$\Gamma = n_{ch}\mu_L F - D_L \partial n_{ch}/\partial x = 0. \tag{4}$$

Here $n_{ch}$ is the density of the protein channels per unit membrane surface; $D_L$ is the lateral diffusion coefficient; $\mu_L$ is the lateral mobility of transmembrane protein channels, which is associated with the lateral diffusion and the local temperature by the Einstein relation:

$$\mu_L = D_L/k_B T. \tag{5}$$

Thus, the steady-state distribution of the protein channels density satisfies to the Boltzmann distribution:

$$n_{ch} = n_0 \exp(-U/k_B T) \tag{6}$$

with the potential

$$U(x) = -\int_0^x F(x)dx. \tag{7}$$

Factor $n_0$ can be easily found from the condition of conservation of the total number of ion channels along the length of the membrane.

$$n_{ch}(x) = \frac{n_{ch,0}}{\frac{1}{l}\int_0^l e^{-U(x)/k_BT}dx} e^{-U(x)/k_BT}, \qquad (8)$$

where $n_{ch,0}$ is the equilibrium density of ion channels without effect of ultrasound. For the case of a standing wave (3):

$$n_{ch}(x) = n_{ch,0} \frac{e^{-\frac{\pi}{3}a^3\overline{E}(1-\cos 2k_n x)/k_BT}}{\frac{1}{l}\int_0^l e^{-\frac{\pi}{3}a^3\overline{E}(1-\cos 2k_n x)/k_BT}dx} = \frac{e^{\frac{\pi}{3}a^3\overline{E}\cos(2k_n x)/k_BT}}{I_o\left(\frac{\pi}{3}\cdot a^3\overline{E}/k_BT\right)}. \qquad (9)$$

Here $I_o$ is modified Bessel function [Korn and Korn, 1968]. The difference between the maximum and minimum densities of ion channels referred to the average density of the channels in the membrane is:

$$\frac{\Delta n_{ch}}{\langle n_{ch}\rangle} = \frac{e^{\frac{\pi}{3}a^3\overline{E}/k_BT}-1}{I_o\left(\frac{\pi}{3}\cdot a^3\overline{E}/k_BT\right)}. \qquad (10)$$

The minimum density of channels for the first harmonic ($n = 1$) is located in the points $x=l/4$ and $3l/4$.

Let us estimate the time $\Delta t_s$ necessary to establish the equilibrium distribution (9) under the influence of ultrasound acoustic pressure force $F$ (1) or (2). The drift velocity of the ion channels is given by: $V_d = \mu_L F$. Hence,

$$\Delta t_s \approx \frac{l}{4V_d} \approx \frac{l\cdot k_BT}{4D_L F_s} \approx \frac{3l^2\cdot k_BT}{8\pi^2 D_L \cdot a^3\overline{E}}. \qquad (11)$$

It is well known that the acoustic waves are damped due to the viscosity as they propagate in an elastic medium,

$$I_s = I_{s,0}e^{-2\alpha x} \qquad (12)$$

with the damping decrement [Mandelstam and Leontovich, 1937]

$$\alpha \approx \frac{1}{2}\frac{\omega^2}{c_s^2\rho}\left(\frac{4}{3}\frac{\eta}{1+\omega^2\tau_\eta^2} + \frac{\mu}{1+\omega^2\tau_\mu^2}\right). \qquad (13)$$

Here $\eta, \mu$, are the shear and bulk viscosities of the medium, and $\tau_\mu$, $\tau_\eta$ are their relaxation times. It is convenient to introduce an effective kinematic viscosity, which we will use in our analysis:

$$v = \frac{v_0}{1+\omega^2\tau_v^2} = \frac{1}{\rho}\left(\frac{4}{3}\frac{\eta}{1+\omega^2\tau_\eta^2} + \frac{\mu}{1+\omega^2\tau_\mu^2}\right). \tag{14}$$

The energy of elastic vibrations in the ultrasonic waves is absorbed due to the viscosity and leads to the heating of medium. The corresponding power of the heat released per unit volume at $\omega\tau_\eta, \omega\tau_\mu \ll 1$ is:

$$W \approx I_{s,0}\alpha \propto I_{s,0}\omega^2. \tag{15}$$

In another extreme case, $\omega\tau_\eta, \omega\tau_\mu \gg 1$, the heat release power is independent on the ultrasound frequency

$$W \approx I_{s,0}\alpha \propto I_{s,0}. \tag{16}$$

It follows from (11) that the time for the transmembrane channels redistribution, under the influence of the force induced by the traveling ultrasonic waves (1), is

$$\Delta t_t \approx \frac{l}{4V_d} \approx \frac{l \cdot k_B T}{4 D_L F_t} \propto \omega^{-4}. \tag{17}$$

Let us estimate the corresponding change in the temperature of the membrane during the time interval (17). For a very high-frequency ultrasound, when $\omega\tau_\eta, \omega\tau_\mu \gg 1$, neglecting the heat removal from the heating release region, and taking into account (16), (17),

$$\Delta T \approx W\Delta t_t / \rho c_p \propto \omega^{-4}. \tag{18}$$

It follows from (1) and (18) that as the frequency of ultrasound increases, the role of the thermal effects decreases, while the rate of the transmembrane channels surface density modification increases, because of $F_t \propto \omega^4$.

Thus, a very high frequency ultrasound can cause redistribution of sodium protein channels and, thereby, stimulate an excitation (or inhibition) of the action potential without significant heat release. This paper shows the principal possibility of generating longitudinal ultrasound in the excitable membrane of myelin nerve fiber as a result of the impact of the low intensity external microwave field, which is insufficient for any noticeable thermal effect. The conditions for the protein transmembrane channels density changes and a possible location where it can happen most effectively were analyzed. The proposed mechanism and the obtained results, in particular, could explain the recent remarkable experiments showing that the ultra-weak, certainly nonthermal, microwave field at specific resonant frequencies may stimulate the activity of nerve fibers (i.e. the initiation of action potentials) [Pikov et al., 2010].

### 3. Statement of the problem

Consider a microwave wave incident on the membrane. For definiteness, let us assume that an electromagnetic wave is plane-polarized, so that the electric field has normal and tangential projections with respect to the membrane surface (Fig. 2). We will consider three characteristic sections of the myelin nerve fibers (see, eg, [Kole and Stuart, 2012; Kole et al, 2008]): the initial segment, the parts covered by the myelin sheath, and the sections separated by the nodes of Ranvier: zones 1, 2 and 3, correspondingly, shown on Fig. 2). The typical lengths of these sections: the initial segment from 10 to 60 μm [Kole and Stuart, 2012]; the nodes of Ranvier, $\sim 1\,\mu m$; the length covered by the myelin sheath, $\sim 100 - 1000\,\mu m$. The typical thickness of the membrane: $d = 2h \approx 6 - 10\,nm$ [Volkenstein, 1983].

Geometrical and mechanical parameters of the nodes of Ranvier, the initial segment and the areas covered by the myelin are very different. Their vibrations (longitudinal and transverse) arising under the influence of a microwave electric field can be considered independently, assuming that the edges of each characteristic section of the axon are clamped. As for the initial segment, since the one end of it rests on the area covered by the myelin layer and the other is in a so-called hillock (see, eg, [Kole et al, 2012]), we will consider it as an abnormally long node of Ranvier. Since the thickness of the axon membrane in the nerve fiber is much smaller than its radius, we will consider the node of Ranvier as well as the initial segment of as a plate, unlimited in the *y*-direction (Fig. 2a-2c), with uniformly distributed surface charge $\sigma_m$, which determines the resting membrane potential. Note, that in the Appendix I it is shown that the problem for these mechanical ultrasound vibrations induced by the interaction with the microwave field in a thin cylindrical membrane is reduced to the vibrations in the flat plate infinite in the normal direction to the surface force.

In general, the direction of the electric field with respect to the axon may be arbitrary. Due to the linearity of the problem, for an arbitrary direction of the wave vector of a plane-polarized microwave field, the mechanical vibrations arising in the membrane are the result of superposition of the vibrations caused by the longitudinal and transverse components of the electric field. Thus, the problem of the impact of microwave on the membrane of the nerve fibers is reduced to the excitation of the ultrasonic standing waves in a plate fixed at the edges, with the given surface charge density. Geometric, mechanical and physical properties of the plate depend on the considered segment of the myelined axon. (The corresponding formulas for the case of the cylindrical membrane and the arbitrary circularly polarized microwave field are shown in the Appendix).

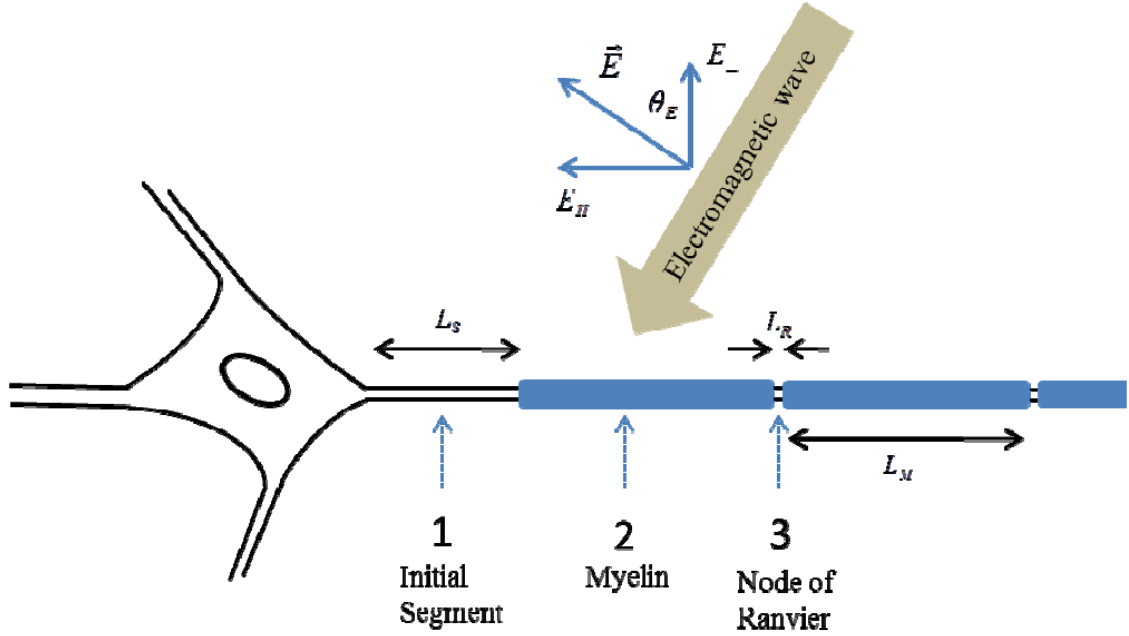

**Figure 2.** Scheme of a neuron with the characteristic segments of the myelined axon interacting with an arbitrary directed plane-polarized microwave field.

The lipid membrane of the axon is usually considered as a capacitor filled with a dielectric with the permittivity $\varepsilon_m \approx 2$ and the surface charge $\sigma_m$. The structure of the ion layer volumetric charge near the surface of the membrane is quite complex, and is described by Gouy-Chapman theory and its modifications [McLaughlin, 1989; Cevc, 1990]. In this paper we assume that the surface charges which form the resting potential are linked to the membrane and not free. Thus, the force with which the electric field acts on these surface charges is directly transferred to the membrane. The corresponding surface charge density can be found from the relationship between the voltage on the capacitor and the charge:

$$\sigma_m = \varepsilon_0 \varepsilon_m V / d , \qquad (19)$$

where $V$ is the potential difference between the capacitor plates (the resting potential). In the unexcited state, the potential difference $V$ between the inner and the outer boundaries of the axon membrane is about -70 mV [Malmivuo J and Plonsey 1995]. Hence, for a typical membrane thickness $d \approx 10$ nm, and the dielectric permittivity we obtain from (19): $\sigma_m \approx 1.23 \cdot 10^{-4}$ C/m$^2$.

It is known that the inner part of the excitable membrane is positively charged relative to the undisturbed electrolyte outside the membrane [McLaughlin, 1989; Cevc, 1990; Franklin et al, 1993; Khalid, 2013]. The case of an interaction of the microwave radiation with the volume charge within the membrane is considered in Appendix II, and it shows that the allowance for internal charge does not lead to qualitative or significant quantitative changes in the results.

The length of the electromagnetic wave in a medium with the dielectric permittivity $\varepsilon$ is [Panofsky and Phillips, 1962]

$$\lambda_{MW} = 2\pi c / \omega_{MW} \sqrt{\varepsilon} . \tag{20}$$

Here, $\omega_{MW} = 2\pi f_{MW}$ is the angular frequency of the wave; $c$ is the speed of light in a vacuum. For the intensity of microwave $I_{MW}$, the corresponding amplitude of the electric field in a medium:

$$E_a = \sqrt{\frac{2I_{MW}}{\varepsilon_0 \varepsilon^{1/2} c}} . \tag{21}$$

In (21), we had taken into account that the intensity remains constant during the transition from vacuum to dielectric medium, i.e.

$$I_{MW} = \frac{\varepsilon_0 \varepsilon E_a^2}{2} c_\varepsilon = \frac{\varepsilon_0 E_{0,a}^2}{2} c , \tag{22}$$

where the phase velocity of light in a medium with the dielectric permittivity $\varepsilon$ is equal to $c_\varepsilon = c / \sqrt{\varepsilon}$; $E_{0,a}$ is the amplitude of the electromagnetic wave in a vacuum. If the ions associated with the surface charge of the resting potential are "below" the membrane surface, then $\varepsilon$ in the expressions (20) and (21) should be set equal to the dielectric permittivity of the membrane, $\varepsilon = \varepsilon_m = 2$. However, if the bound ions are "above" the surface of the membrane, then $\varepsilon = \varepsilon_L(\omega_{MW})$. For the microwave radiation in water in the frequency range 50-300 GHz, $\varepsilon_L(\omega_{MW}) \approx 5$ [Buchner et al, 1999]. Taking into account that the dependence of the field amplitude on the dielectric permittivity (21), $E_a \sim \varepsilon^{-1/4}$, all the results and conclusions will remain practically unchanged.

In the microwave range from hundreds of megahertz to hundreds of gigahertz, the wavelength of the electromagnetic radiation is considerably larger than any sections of our interest in the myelin nerve fiber (Fig. 2). Therefore, the external electric field interacting with these regions can be considered uniform, and only depending on time. The direction of the electric field relative to the axon is not fixed, it can be arbitrary. So, as we have already mentioned, we will consider the two possible cases: when the electric field is directed parallel to the surface and when it is along the normal to the membrane (Fig. 2).

In accordance with the statement of the problem, each section of the axon that belongs to the node of Ranvier or to the initial segment, is considered as a plate, unlimited in the *y* direction and clamped at the edges (Fig. 3). The figures 3 correspond to any of the sections of the myelin nerve fibers (node of Ranvier or to the initial segment) in which we are studying the forced vibrations induced by the microwave. Since, as shown in the previous section, the ions responsible for the resting potential are bonded by the electrostatic forces, the electric field perpendicular to the plate will buckle the plate (Fig. 3a-3b), and the parallel will result in the shear deformation (Fig. 3c-3d ). We believe that in a weak microwave field, the free longitudinal displacement of the ions, which are responsible for the resting potential, is limited along the surface of the membrane, since, as shown above, they are not free. Thus, their displacement due to the influence of the external field will be transferred to the membrane.

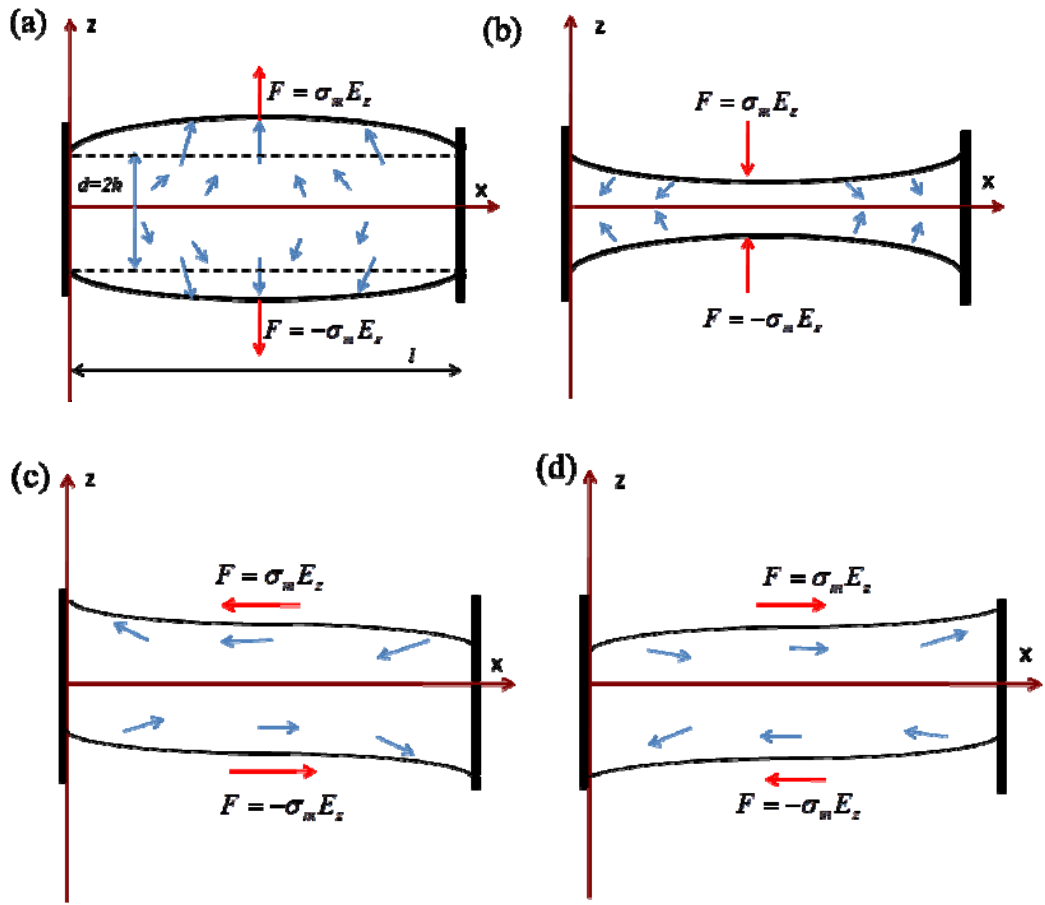

**Figure 3.** Membrane deformations causes by the microwave electric field directed perpendicular ((a) and (b)) and parallel ((c) and (d)) to the surface. The edges of the membrane (unlimited in *y* direction) are clamped. The blue arrows indicate the elastic displacements inside the membrane. It can be seen that even a force normal to the membrane causes deformations with longitudinal component.

These types of displacements are well-studied for the forced deformations of an elastic bar with clamped ends in the classic theory of elasticity [Landau and Lifshitz, 1970].

## 4. Model of the membrane vibrations in the microwave electromagnetic field

The equations describing the longitudinal and transverse vibrations of a membrane, such an elastic bar under the influence of external driving forces, have the form [Landau and Lifshitz, 1970]:

$$\frac{\partial^2 \mathbf{u}}{\partial t^2} = c_t^2 \Delta \mathbf{u} + (c_l^2 - c_t^2)\nabla(\nabla \cdot \mathbf{u})$$
$$\mathbf{u} = \mathbf{i}_x u_x + \mathbf{i}_y u_y + \mathbf{i}_z u_z$$
(23)

Here $\mathbf{u}$ is the displacement vector; $c_l, c_t$ are the longitudinal (along the *x*-axis) and transverse (along the *z*-axis) speed of sound. In our model, we assume that $c_l \approx c_t$. For a planar plate of infinite extent in the y direction, equation (23), written for the displacement projections takes the form

$$\frac{\partial^2 u_x}{\partial t^2} = c_l^2 \frac{\partial^2 u_x}{\partial x^2} + c_t^2 \frac{\partial^2 u_x}{\partial z^2} + \nu \frac{\partial}{\partial t}\left(\frac{\partial^2 u_x}{\partial x^2}\right) + (c_l^2 - c_t^2)\frac{\partial^2 u_z}{\partial z \partial x}$$
$$\frac{\partial^2 u_z}{\partial t^2} = c_t^2 \frac{\partial^2 u_z}{\partial x^2} + c_l^2 \frac{\partial^2 u_z}{\partial z^2} + \nu \frac{\partial}{\partial t}\left(\frac{\partial^2 u_z}{\partial x^2}\right) + (c_l^2 - c_t^2)\frac{\partial^2 u_x}{\partial z \partial x}$$
(24)

where longitudinal dissipative friction terms have been included. Here $\nu$ is the effective kinematic viscosity (14) related to the dissipation of the induced vibrations of the membrane due to the frictions on water.

The boundary conditions corresponding to the electric field perpendicular to the surface of the membrane (Fig. 3a,3b) are:

$$\left.\frac{\partial u_z}{\partial z}\right|_{z=h} = \frac{F_z}{c_l^2 \rho_m}$$
$$\left.u_z\right|_{z=0} = 0$$
$$\left.u_z\right|_{x=0, x=L} = 0$$
(25)

Here $\rho_m$ is the density of the membrane; $F_z$ is the normal to the surface force acting on the membrane by the microwave electric field

$$F_z = \sigma_m E_\perp = \sigma_m \sqrt{\frac{2I_{MW}}{\varepsilon_0 \varepsilon_m^{1/2} c}} e^{i\omega t} \cos(\theta_E). \tag{26}$$

Where $\sigma_m$ is the surface charge density on the membrane; $\theta_E$ is the angle between the direction of the electric field and the *z*-axis (Fig.2).

The boundary conditions corresponding to the longitudinal electric field (Fig. 3c,3d) are:

$$\left.\frac{\partial u_x}{\partial z}\right|_{z=h} = \frac{F_x}{c_l^2 \rho_m}$$
$$u_x|_{z=0} = 0 \tag{27}$$
$$u_x|_{x=0, x=L} = 0$$

Here $F_x$ is the tangential to the surface force

$$F_x = \sigma_m E_{II} = \sigma_m \sqrt{\frac{2I_{MW}}{\varepsilon_0 \varepsilon_m^{1/2} c}} e^{i\omega t} \sin(\theta_E). \tag{28}$$

For definiteness, we assume that the electric field lies in the *xz* plane.

### 4.1. The electric field perpendicular to the membrane surface

We will look for a solution of equations (24), in accordance with the boundary conditions (25), in the form

$$u_x = A_{II} e^{i\omega t} \cos(k_n x)\cos(\xi z)$$
$$u_z = A_\perp e^{i\omega t} \sin(k_n x)\sin(\xi z) \tag{29}$$
$$k_n = \frac{\pi n}{L}$$

Substituting (29) into (24), we obtain:

$$(\omega^2 - k_n^2 c_l^2 - i\omega v k_n^2 - c_t^2 \xi^2)A_{II} + k_n(c_l^2 - c_t^2)\xi A_\perp = 0$$
$$(\omega^2 - k_n^2 c_t^2 - i\omega v k_n^2 - c_l^2 \xi^2)A_\perp + k_n(c_l^2 - c_t^2)\xi A_{II} = 0 \tag{30}$$

$$c_l^2 \xi \cos(\xi h) \cdot A_\perp = \frac{F_{n,z}}{\rho} \tag{31}$$

where $F_{n,z}$ is the *n*-th harmonic of the normal force.

The dispersion equation follows from two first equations (30):

$$(\omega^2 - k_n^2 c_l^2 - i\omega v k_n^2 - c_t^2 \xi^2) \cdot (\omega^2 - k_n^2 c_t^2 - i\omega v k_n^2 - c_l^2 \xi^2) = k_n^2 (c_l^2 - c_t^2)^2 \xi^2, \tag{32}$$

and links the wave number of the transverse displacement along *z*-axis with the frequency of the external electric field and the longitudinal wave vector $k_n$. It follows from equations (30) and (31) that the amplitudes of the transverse $A_\perp$ and longitudinal $A_{II}$ displacements are

$$A_\perp = \frac{1}{c_l^2 \xi \cos(\xi h)} \frac{F_{n,z}}{\rho}, \tag{33}$$

$$A_{II} = -\frac{k_n (c_l^2 - c_t^2) \xi}{\omega^2 - k_n^2 c_l^2 - i\omega v k_n^2 - c_t^2 \xi^2} \cdot \frac{1}{c_l^2 \xi \cos(\xi h)} \frac{F_{n,z}}{\rho}. \tag{34}$$

Since we are interested in the frequency range $\omega \gg k_n c_l, k_n c_t$, then it follows from (32)

$$\xi^2 \approx \frac{\omega^2 - i\omega v k_n^2}{c_l^2}. \tag{35}$$

Substituting (35) into (33) and to (34), we obtain:

$$A_\perp = \frac{1}{c_l \omega \cos(\xi h)} \frac{F_{n,z}}{\rho}$$

$$A_{II} = -\frac{k_n}{\omega^2} \frac{1}{\cos(\xi h)} \frac{F_{n,z}}{\rho} \tag{39}$$

Taking into account that

$$|\cos(\xi h)| = \left|\cos\left(\frac{\omega h}{c_l} + i\frac{vk^2 h}{2c_l}\right)\right| = \sqrt{\cos^2\left(\frac{\omega h}{c_l}\right) + sh^2\left(\frac{vk^2 h}{2c_l}\right)}, \tag{37}$$

we can find from (35) and (36) the frequencies and the widths of resonances.

Since $\frac{vk^2 h}{2c_l} \ll 1$, the resonant cyclic frequencies are

$$\omega_r = \left(\frac{\pi}{2} + \pi p\right)\frac{c_l}{h} \ll 1, \quad p = 0,1,2,\ldots, \tag{38}$$

and the width of the resonances is:

$$\delta\omega \approx vk_n^2 / 2. \tag{39}$$

Substituting $\omega_r$ in (36), we find the resonant amplitudes $A_\perp$ and $A_{II}$

$$A_{r,\perp} = \frac{2}{\omega_r \nu k_n^2 h} \frac{F_{n,z}}{\rho_m}$$

$$A_{r,II} = -\frac{c_l k_n}{\omega_r} A_{r,\perp}$$
(40)

The force $F_{n,z}$ in (40) is defined by (26).

### 4.2. The electric field parallel to the surface of the membrane

In another extreme case, there are forced mechanical vibrations under the influence of an electric field parallel to the plane of surface. In accordance with the boundary conditions (27), will be looking for the solution of (24) in the form

$$u_x = A_{II} e^{i\omega t} \sin(k_n x)\sin(\xi z)$$
$$u_z = A_\perp e^{i\omega t} \cos(k_n x)\cos(\xi z)$$
$$k_n = \frac{\pi n}{L}$$
(41)

Substituting (46) into (27) and (30) we obtain

$$\left(\omega^2 - k_n^2 c_l^2 - i\omega\nu k_n^2 - c_t^2 \xi^2\right)A_{II} + k_n \left(c_l^2 - c_t^2\right)\xi A_\perp = 0$$
$$\left(\omega^2 - k_n^2 c_t^2 - i\omega\nu k_n^2 - c_l^2 \xi^2\right)A_\perp + k_n \left(c_l^2 - c_t^2\right)\xi A_{II} = 0$$
(42)

$$c_t^2 \xi \cos(\xi h) \cdot A_{II} = \frac{F_{n,x}}{\rho}$$
(43)

Here $F_{n,x}$ is the $n$th harmonic of the longitudinal force. Expressions (42) and (43), up to replacing $A_{II}$ by $A_\perp$, $F_{n,x}$ by $F_{n,z}$, and $c_l$ by $c_t$, coincide with the (30), (31).

### 5. Results and Discussion

All quantitative estimates presented in this paper are obtained for the following set of parameters typical for the lipid membrane of the nerve fiber: $\rho_m \approx 900 \text{kg/m}^3$, $d = 2h = 10$ nm; $c_l \approx c_t \approx 1500 \text{m/s}, \varepsilon_m = 2; \sigma_m \approx 1.24 \cdot 10^{-4}$ C/m$^2$, corresponding to the resting membrane potential of $\approx 70$ mV. The kinematic viscosity $\nu$ is given by (14). For water at 20 degrees Celsius: $\nu_0 \approx 10^{-6}$ m$^2$/s, $\tau_\nu \approx 8 \cdot 10^{-12}$ s [Nimtz and Weiss, 1987)].

From (38) it follows that the first resonance frequency is $f_1 = 74.6$ GHz. Accordingly, the kinematic viscosity (14) for the first resonance frequency is $\nu = 6.3 \cdot 10^{-8}$ m$^2$/s.

Subsequent resonances are separated by quite large frequency intervals. Thus, the second resonance frequency is $f_2 = 223.9\,\text{GHz}$, the third is $f_3 = 371.3\,\text{GHz}$, and so on. Since the amplitudes of the forced oscillations (40) are inversely proportional to the production $\omega_r \nu$, and the viscosity at high frequencies $\nu \propto 1/\omega^2$ (13), the resonant amplitude increases linearly with frequency. However, the products $A_{r,\perp}\delta\omega, A_{r,\text{\it II}}\delta\omega \propto 1/\omega_r$ decrease linearly with the frequency.

It should be noted that due to the existence of natural statistical variations of geometrical parameters of the nerve fiber regions - their length and the membrane thickness - the response of the ensemble of neurons on the microwave will have a much broader spectral range of in the vicinity resonances. Let us estimate a spectral width of the resonance for the initial segment. From (29) it follows:

$$\omega^2 = c_t^2 \xi^2 + k^2 c_l^2. \tag{44}$$

Aт the first resonance, $\xi = \dfrac{\pi}{2h}$. Then, at $c_t \approx c_l$, it follows from the dispersion equation (32)

$$\omega_r^2 = \left(\frac{\pi}{2}\frac{c_l}{h}\right)^2 + k_n^2 c_l^2. \tag{45}$$

Hence, for $k_1 = \dfrac{\pi}{l}$ we get:

$$f_r = \frac{\omega_r}{2\pi} = \frac{c_l}{4h}\left(1 + \frac{2h^2}{l^2}\right). \tag{46}$$

Since the length of the processes are varying between 10-50 μm, then $\Delta f_r \approx 22\,\text{kHz}$. In fact, the resonance can be even more broadened if we consider subharmonics $k_n = n\pi/l$, and also that the thickness of the membrane may vary within the limits $6-10$ nm.

It is interesting that our simple analysis leads to the resonance microwave frequencies, which are close enough to those which have been experimentally observed [Pikov et al, 2010]. Note that the excitation of mechanical vibrations in charged biological membranes as a result of interaction with a weak microwave field and the its resonant nature were also theoretically studied in [Krasilnikov, 1999; 2001].

A critical moment of our analysis is the density distribution of ion channels and the time of its establishment. On the basis of formulas (2), (10), knowing the dependence of the resonance amplitude of the longitudinal vibrations on the intensity of the incident electromagnetic wave and the surface charge density on the membrane, it is easy to find the ratio $U/k_B T$ and to estimate the time $\Delta t_s$ for establishment of an equilibrium distribution of the sodium channels density in the initial segment or the nodes of Ranvier. The greatest contribution to the amplitude of the forced longitudinal vibrations of the membrane gives the longitudinal component of the electric field, for which the results are shown in Table 1. The amplitude of the displacement at the longitudinal electric field is inversely proportional to the resonant frequency:

$$A_{r,\parallel} = \frac{2}{\omega_r v k^2 h} \frac{F_{n,x}}{\rho_m}. \qquad (47)$$

The expression for the force acting on the ion channel follows from (3) and (47), when it is assumed in (3) that $a = h = d/2$:

$$F = \frac{8\pi}{3} \frac{h}{k_n^3 v^2} \sigma_m^2 \frac{I_{MW}}{\rho \varepsilon_0 \varepsilon_m^{1/2} c} \sin(2k_n x) \qquad (48)$$

The time for establishment of the equilibrium distribution of ionic channels and relative density change can be calculated using (48), (10) and (11). Table 1 shows the results for the longitudinal vector of the fundamental mode $k = \pi/l$, where $l$ is the length of the nodes of Ranvier ($l = 2.5\,\mu m$) or the initial segment ($l = 15$, $30$, $50\,\mu m$). In all presented variants the coefficient of lateral diffusion was assumed to be $D_L = 10^{-13}$ m$^2$/s, which is typical for transmembrane protein Na$^+$ channels in the nerve fiber outside the hillock (region connecting the neuron with the initial segment) [Jacobson, 1987]. This value of lateral diffusion is a lower limit and it significantly increases with increasing temperature (see for example [Petrov and Schwille, 2008]). This process accelerates the reaction of the nervous fiber to the microwave radiation, but in the present work we will not stop on the detailed analysis of this interesting problem .

As seen from the table, the periodic mechanical oscillations caused by the longitudinal resonance electric field, result in a noticeable density redistribution of protein sodium channels in the initial segment (Fig. 2), even when the intensity of the microwave is only about 1 W/m$^2$. Note that a significant modification of the action potential threshold is enough to change the density of sodium channels by only 20-30% [Platkiewicz and Brette, 2010], so the response time of the nerve fibers to the microwave can be significantly lower than specified in the table.

As expected, the greatest change in density of sodium channels per unit surface area of the membrane occurs in the longest initial segments (see Fig.2). In nodes of Ranvier stimulated change in the density of channels is too small to affect the threshold of the action potential.

**Table 1.** The results of calculations for the nodes of Ranvier and for the the initial segment at various characteristic lengths for the normal incidence of the microwave on the membrane.

| $l$ [μm] | $I_{MW}$ [W/m$^2$] | $U/k_B T$ | $\Delta n_{ch}/\langle n_{ch}\rangle$ | $\Delta t$ [minutes] |
|---|---|---|---|---|
| 2.5 | 0.01 | 2.33E-08 | 2.33E-08 | 1.78E+06 |
| 15  | 0.01 | 3.01E-05 | 3.01E-05 | 4.94E+04 |
| 30  | 0.01 | 4.82E-04 | 4.82E-04 | 1.24E+04 |
| 50  | 0.01 | 3.73E-03 | 3.74E-03 | 4.45E+03 |
|     |      |          |          |          |
| 2.5 | 0.1  | 2.33E-07 | 2.33E-07 | 1.78E+05 |
| 15  | 0.1  | 3.01E-04 | 3.01E-04 | 4.94E+03 |
| 30  | 0.1  | 4.85E-03 | 4.86E-03 | 1.24E+03 |
| 50  | 0.1  | 3.72E-02 | 3.79E-02 | 4.45E+02 |
|     |      |          |          |          |
| 2.5 | 1    | 2.33E-06 | 2.33E-06 | 1.78E+04 |
| 15  | 1    | 3.04E-03 | 3.04E-03 | 4.94E+02 |
| 30  | 1    | 4.82E-02 | 4.94E-02 | 1.24E+02 |
| 50  | 1    | 3.72E-01 | 4.35E-01 | 4.45E+01 |
|     |      |          |          |          |
| 2.5 | 10   | 2.33E-05 | 2.33E-05 | 1.78E+03 |
| 15  | 10   | 3.01E-02 | 3.06E-02 | 4.94E+01 |
| 30  | 10   | 4.82E-01 | 5.85E-01 | 1.24E+01 |
| 50  | 10   | 3.72E+00 | 4.53E+00 | 4.45E+00 |
|     |      |          |          |          |
| 2.5 | 100  | 2.33E-04 | 2.33E-04 | 1.78E+02 |
| 15  | 100  | 3.01E-01 | 3.44E-01 | 4.94E+00 |
| 30  | 100  | 4.82E+00 | 5.30E+00 | 1.24E+00 |
| 50  | 100  | 3.72E+01 | 1.52E+01 | 4.45E-01 |

In this work we have shown that there is a fundamental possibility of generation of the longitudinal standing ultrasonic waves in excitable membrane of the myelinend nerve fiber by a low-intensity external microwave field at frequencies of 50 - 300 GHz, which is insufficient for any significant thermal effect. Also, we found the dependences of changes in the density of protein transmembrane sodium channels on the membrane thickness and length, and the intensity of the microwave field. The most favorable region for spontaneous action potential occurrence is the initial segment. In particular, obtained results and analysis may explain amazing recent experiments which have shown that the ultra-weak, certainly non-thermal microwave fields at specific resonant frequencies could stimulate the activity of the nerve fibers (by initiation of the action potentials) [Pikov et al, 2010].

**Conclusions**

1. It is shown that the non-thermal microwave field can cause the ultrasonic mechanical vibrations in the nerve fibers.
2. The presence of the resonances in the range of tens of GHz is in good agreement with known experimental data.
3. The interaction with the ultrasound vibrations could change the density of transmembrane protein sodium channels. As a result, it could also change the resting potential in the nerve fibers and, therefore, it may induce or, conversely, suppress the excitation of the action potentials.
4. The most effective component of the microwave electric field is the tangential component to the surface of the membrane, and the most effective region in the myelined axon is the initial segment, i.e., the section between the neuron and the first portion covered with a myelin sheath.
5. Despite the encouraging results, our analysis is preliminary and requires further in-depth experimental and theoretical study. First of all, this analysis concerns with issues related to the density of the surface charges, the longitudinal and transverse sound velocities of in the membrane and the surrounding fluid, and viscous losses at GHz frequencies range, where the amplitude of ultrasonic vibrations is comparable to, or less than, the intermolecular distance in the liquid and membrane.
6. In this work we considered only the influence of ultrasonic standing waves excited by microwave radiation on the distribution of $Na^+$ transmembrane protein channels. However, similar effects should be expected for other kinds of membrane proteins (such as transmembrane ion channels, as well as the peripheral surface proteins) affected by lateral diffusion. This, in principle, could significantly increase the effects caused in the nerve fibers by the weak microwave radiation.

**Appendix I. Elastic cylindrical membrane.**

We considered a model membrane as a plain elastic bar with clamped edges under the influence of longitudinal or normal forces. The actual nerve fiber rather is a tube formed by the lipid membrane (Fig. 4). We will show that all of the above qualitative and quantitative results remain valid.

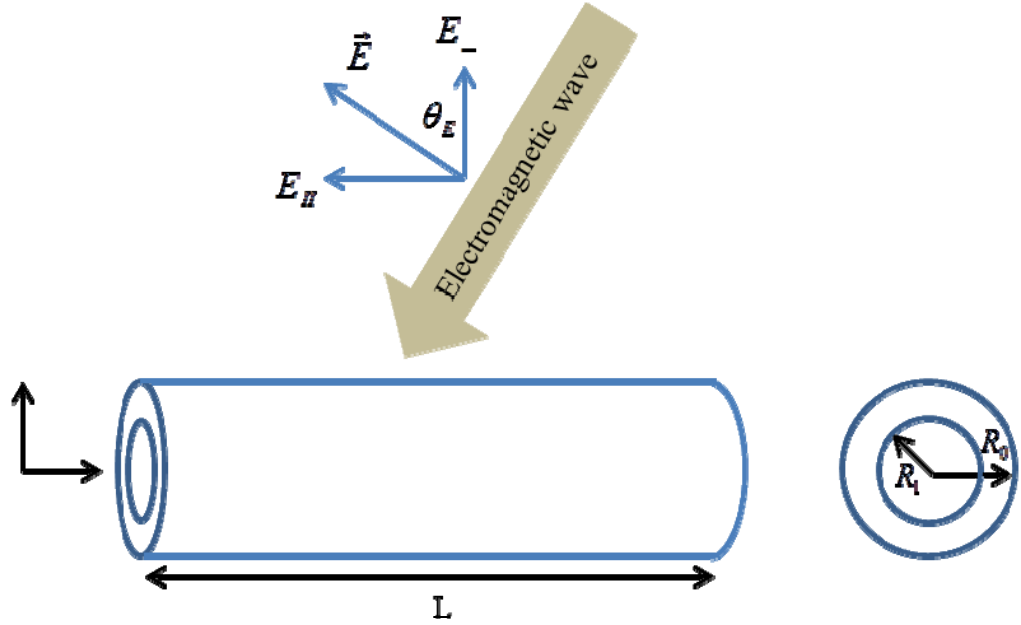

**Figure 4.** Elastic cylindrical membrane of the nerve fibers in the microwave field. The outer surface of the cylinder is negatively charged, whereas the internal surface is positive.

Because the main contribution to the redistribution of the channels along myelinated fibers results in forces directed along the fiber, we restrict ourselves only to this case. The equations of the theory of elasticity (2.1) in a cylindrical coordinate system, taking into account the friction of the surface membrane on the water, have the form [Landau and Lifshitz, 1970].

$$\frac{\partial^2 u_r}{\partial t^2} = c_t^2 \left( \frac{\partial^2 u_r}{\partial z^2} - \frac{\partial^2 u_z}{\partial r \partial z} \right) + c_l^2 \left( \frac{\partial^2 u_r}{\partial r^2} + \frac{1}{r}\frac{\partial u_r}{\partial r} - \frac{u_r}{r^2} + \frac{\partial^2 u_z}{\partial r \partial z} \right) + v \frac{\partial}{\partial t} \frac{\partial^2 u_r}{\partial z^2}$$

(A.1)

$$\frac{\partial^2 u_z}{\partial t^2} = c_t^2 \left( \frac{1}{r}\frac{\partial u_z}{\partial r} + \frac{\partial^2 u_z}{\partial r^2} - \frac{\partial^2 u_r}{\partial r \partial z} - \frac{1}{r}\frac{\partial u_r}{\partial z} \right) + c_l^2 \left( \frac{\partial^2 u_r}{\partial r \partial z} + \frac{1}{r}\frac{\partial u_r}{\partial z} + \frac{\partial^2 u_z}{\partial z^2} \right) + v \frac{\partial}{\partial t} \frac{\partial^2 u_z}{\partial z^2}$$

For a cylinder with clamped edges, when the surface forces oscillating with the frequency $\omega$, solutions of equations (A.1) have the form:

$$u_r = U_{n,r} e^{i\omega t} \cos(k_n z)$$
$$u_z = U_{n,z} e^{i\omega t} \sin(k_n z)$$

(A.2)

The corresponding boundary conditions [Landau and Lifshitz, 1970]

$$c_t^2 \rho \left( \frac{\partial U_{n,r}}{\partial z} + \frac{\partial U_{n,z}}{\partial r} \right)_{r=R_0} = F_{\|,n,R_0}$$

$$c_t^2 \rho \left( \frac{\partial U_{n,r}}{\partial z} + \frac{\partial U_{n,z}}{\partial r} \right)_{r=R_1} = F_{\|,n,R_1}$$

(A.3)

where

$$U_{n,r} = A_{n,r,J} J_1(\xi r) + A_{n,r,Y} Y_1(\xi r)$$
$$U_{n,z} = A_{n,z,J} J_0(\xi r) + A_{n,r,Y} Y_0(\xi r).$$

(A.4)

Here $J_1$, $Y_1$ are the Bessel functions; $k_n = \pi n / L$; $R_0$, $R_1$ are the external and the internal radiuses of the cylinder; $F_{\|,n,R_0}$, $F_{\|,n,R_1}$ are the amplitudes of the surface forces. For microwave field with the electric field parallel to the cylinder axis, $F_{\|,n,R_1} = -F_{\|,n,R_0}$.

Substituting into (A.1) and (A.3) expressions (A.4) for $U_{n,z}$ and $U_{n,r}$, we obtain:

$$\left( (\omega^2 - c_t^2 k_n^2 - i\omega\nu k_n^2) - c_l^2 \xi^2 \right) \cdot A_{n,r,j} + (c_t^2 - c_l^2) \cdot k_n \xi \cdot A_{n,z,j} = 0$$
$$\left( (\omega^2 - c_t^2 k_n^2 - i\omega\nu k_n^2) - c_l^2 \xi^2 \right) \cdot A_{n,z,j} + (c_t^2 - c_l^2) \cdot k_n \xi \cdot A_{n,r,j} = 0$$
$$\left( (\omega^2 - c_t^2 k_n^2 - i\omega\nu k_n^2) - c_l^2 \xi^2 \right) \cdot A_{n,r,Y} + (c_t^2 - c_l^2) \cdot k_n \xi \cdot A_{n,z,Y} = 0$$
$$\left( (\omega^2 - c_t^2 k_n^2 - i\omega\nu k_n^2) - c_l^2 \xi^2 \right) \cdot A_{n,z,Y} + (c_t^2 - c_l^2) \cdot k_n \xi \cdot A_{n,r,Y} = 0$$

(A.5)

The boundary conditions (A.3) reduce to

$$\left( -k_n A_{n,r,J} - \xi A_{n,z,J} \right) J_1(\xi R_0) + \left( -k_n A_{n,r,Y} - \xi A_{n,z,Y} \right) Y_1(\xi R_0) = \frac{F_{\|,n,R_0}}{c_t^2 \rho}$$

$$\left( -k_n A_{n,r,J} - \xi A_{n,z,J} \right) J_1(\xi R_1) + \left( -k_n A_{n,r,Y} - \xi A_{n,z,Y} \right) Y_1(\xi R_1) = \frac{F_{\|,n,R_0}}{c_t^2 \rho}$$

(A.6)

Here we have used known relations for Bessel functions [Korn and Korn, 1968]

$$J_1(\xi x) = -\frac{1}{\xi} \frac{dJ_0(\xi x)}{dx}$$

$$J_0(\xi x) = \frac{1}{\xi x} J_1(\xi x) + \frac{1}{\xi} \frac{dJ_1(\xi x)}{dx}$$

From (A.5) the dispersion equation follows:

$$\left( (\omega^2 - c_t^2 k_n^2 - i\omega\nu k_n^2) - c_l^2 \xi^2 \right) \cdot \left( (\omega^2 - c_t^2 k_n^2 - i\omega\nu k_n^2) - c_t^2 \xi^2 \right) = (c_t^2 - c_l^2)^2 \cdot k_n^2 \xi^2.$$

(A.7)

Equation (A.7) coincides with (35). The system of equations (A.5) and (A.6) gives the complete solution to the problem of cylinder (membrane) vibrations under periodic longitudinal surface axially symmetric forces. For a thin-walled cylinder ($2h = R_0 - R_1 \ll R_0 \ll L$) with clamped edges, the problem of elastic vibrations under the action of the longitudinal forces coincides with the problem of the plain plate vibrations considered in section 4.

**Appendix II. Longitudinal vibrations of the charged membrane**

In Part 3 (Statement of Problem), we assumed that the membrane of the nerve fiber is a parallel-plate capacitor charged to the resting potential with the charges on the plates tightly bound to the membrane surface. In fact, the volume of the membrane can be positively charged [McLaughlin, 1989; Cevc, 1990; Franklin et al, 1993; Khalid, 2013]. In this case, for the unexcited membrane the potential difference between the center of the membrane and the unperturbed electrolyte at infinity is equal to the so-called dipole potential $V_d \approx 200\,\text{mV}$. Let us show that taking into account the dipole potential of all the results obtained above remains valid. However, in the case of a volume charge, both surfaces of the membrane are vibrating in the same direction unlike shown in Fig. 3c-3d.

Equations (24) with the body force directed along the membrane take the form:

$$\frac{\partial^2 u_x}{\partial t^2} = c_l^2 \frac{\partial^2 u_x}{\partial x^2} + c_t^2 \frac{\partial^2 u_x}{\partial z^2} + v \frac{\partial}{\partial t}\left(\frac{\partial^2 u_x}{\partial x^2}\right) + \left(c_l^2 - c_t^2\right)\frac{\partial^2 u_z}{\partial z \partial x} + \frac{F_{vol}}{\rho_m}$$

(A2.1)

$$\frac{\partial^2 u_z}{\partial t^2} = c_t^2 \frac{\partial^2 u_z}{\partial x^2} + c_l^2 \frac{\partial^2 u_z}{\partial z^2} + v \frac{\partial}{\partial t}\left(\frac{\partial^2 u_z}{\partial x^2}\right) + \left(c_l^2 - c_t^2\right)\frac{\partial^2 u_x}{\partial z \partial x}$$

where $F_{vol} = E_x q(z)$; $q$ is the volume charge density.

Assuming that the volume charge uniformly distributed within the membrane and the potential difference between the center of the membrane and its edges is equal to the dipole potential $V_d$, we can find the potential distribution inside the membrane $\varphi = \frac{q}{2\varepsilon\varepsilon_0} z^2$. And, accordingly, $q = 2\varepsilon\varepsilon_0 \frac{V_d}{h^2} \approx 2.83 \cdot 10^5\,C/m^3$.

We will look for the solution of (A2.1) in the form:

$$u_x = A_{II}(z)e^{i\omega t}\sin(k_n x)$$
$$u_z = A_\perp(z)e^{i\omega t}\cos(k_n x) \qquad (A2.2)$$
$$k_n = \frac{\pi n}{l}$$

Substituting (A2.2) for (A2.1), we get:

$$-\omega^2 A_{II} = -c_l^2 k_n^2 A_{II} + c_t^2 \frac{\partial^2 A_{II}}{\partial z^2} + i\omega v k_n^2 A_{II} - ik_n(c_l^2 - c_t^2)\frac{\partial A_\perp}{\partial z} + \frac{F_{vol}}{\rho_m}$$
$$-\omega^2 A_\perp = -c_t^2 k_n^2 A_\perp + c_l^2 \frac{\partial^2 A_\perp}{\partial z^2} + i\omega v k_n^2 A_\perp + ik_n(c_l^2 - c_t^2)\frac{\partial A_{II}}{\partial z} \qquad (A2.3)$$

In the case of high frequencies, it follows from (A2.3) for longitudinal oscillations:

$$\left(-\frac{\omega^2}{c_t^2} - i\frac{\omega v k_n^2}{c_t^2}\right)A_{II} = \frac{\partial^2 A_{II}}{\partial z^2} + \frac{F_{vol}}{c_t^2 \rho_m} \qquad (A2.4)$$

The boundary conditions for (A2.3), assuming that the whole charge is located inside the membrane, have the form:

$$\left.\frac{\partial A_{II}}{\partial z}\right|_{z=h} = 0 \qquad \left.\frac{\partial A_{II}}{\partial z}\right|_{z=-h} = 0 \qquad (A2.5)$$

It is convenient to present the longitudinal displacement and the bulk force in the form:
$$A_{II} = \sum_{p=0,1,2,\ldots} A_{II,p}\cos(k_p z), \quad F_{vol} = \sum_{p=0,1,2,\ldots} F_{vol,p}\cos(k_p z) \qquad (A2.6)$$

Where,

$$k_p = \left(\frac{1}{2} + p\right)\frac{\pi}{h}, \qquad (A2.7)$$

and

$$A_{II,p} = \frac{1}{(k_p^2 c_t^2 - \omega^2) - i\omega v k_n^2}\frac{F_{vol,p}}{\rho_m} \qquad (A2.8)$$

Hence, the resonant frequencies:

$$f_r = \frac{\omega_r}{2\pi} = \left(\frac{1}{2} + p\right)\frac{c_t}{2h} = \left(\frac{1}{2} + p\right)\frac{c_t}{d} \qquad (A2.9)$$

and amplitudes,

$$A_{II,r,p} = \frac{1}{\omega_r v k_n^2} \frac{F_{vol,p}}{\rho_m} = \frac{2}{\omega_r v k_n^2 h} \frac{hF_{vol,p}}{2\rho_m} \tag{A2.10}$$

As expected, the resonant frequencies (A2.9) and (38) coincide with each other. The corresponding "effective" surface charge density $\frac{hF_{vol,p}}{2} \approx 7.1 \cdot 10^{-4} \frac{C}{m^2}$ is 5.7 times greater than the surface charge density $\sigma_m$ used in our calculations. This means that the corresponding time $\Delta t_s$ for protein sodium channels adjustment is about 30 times less than shown in Table 1.